\documentstyle[sprocl]{article}
\def\Journal#1#2#3#4{{#1} {\bf #2}, #3 (#4)}

\def\be{\begin{equation}}
\def\ee{\end{equation}}
\def\bea{\begin{eqnarray}}
\def\eea{\end{eqnarray}}

\title{
  \begin{flushright} {\small $\begin{array}{ l } \mbox{HD--THEP--96--57} \\
    \mbox{WUB--96--40} \end{array} $}
 \end{flushright}
\vskip 1cm
COMPUTER AIDED SERIES EXPANSIONS FOR CRITICAL PHENOMENA \footnote 
{Talk presented by H.~Meyer-Ortmanns at the International Conference
  on MULTI-\\
 SCALE PHENOMENA AND THEIR SIMULATION at ZiF (Bielefeld) October 1996}}
\author{HILDEGARD MEYER-ORTMANNS \footnote
 {e-mail:~ortmanns@thphys.uni-heidelberg.de}}

\address{Theoretical Physics, University of Wuppertal\\ D-42097
  Wuppertal, Germany} 

\author{THOMAS REISZ \footnote{e-mail:~reisz@thphys.uni-heidelberg.de}}

\address{Theoretical Physics, University of Heidelberg\\ 
 D-69121 Heidelberg, Germany}

\begin{document}

\maketitle

\abstracts{
Under quite general conditions
critical phenomena can be described with high order linked cluster
expansions. 
The coefficients of the series admit a graphical expansion that
is generated with the aid of computers. Our generalization of linked
cluster expansions from an infinite to a finite volume allows
to perform a finite size scaling analysis. 
We also indicate a generalization to {\it Dynamical Linked Cluster 
Expansions} with possible applications to spin glasses and neural
networks with coupled spin and interaction dynamics.}
  

\section{Linked Cluster Expansions in the Infinite Volume}

We will focus on multiscale phenomena that occur as {\it critical} phenomena 
in statistical systems at second order phase transitions. We consider 
statistical systems with regions of first and  second order transition regions
in phase space. Our computational tools are analytical rather than numerical
calculations. We use convergent series expansions in a parameter called 
hopping parameter 
$\kappa$. When applied to a calculation of the free energy and 
connected correlations, the expansions amount to Linked Cluster Expansions 
(LCEs). The expansion coefficients admit a representation as a sum of 
connected graphs.
The price one has to pay for describing critical phenomena with an accurracy of
the order of
$1 \%$ is a high order in the expansion parameter. Since the number of 
contributing graphs rapidly increases with increasing order in $\kappa$, a
computer aided algorithmic generation of graphs becomes unavoidable. Even an
optimization of algorithms becomes essential to make the handling of 
millions of graphs feasible. Such a large number of graphs has to be
dealt with to compute two and four-point functions to 20th order in 
$\kappa$ in LCEs. 
For example, this order is taken into account in scalar $O(N)$
models at finite temperature in order
to measure the tiny finite temperature effects on the phase structure 
\cite{reisz}.

Originally, linked cluster expansions have been developed in the infinite 
volume. The reason why we have generalized them to a finite volume was twofold.
One reason was to identify 1st order transitions from series
expansions in the high temperature phase. 
The second reason was to distinguish second order transitions
associated with different universality classes. 


\subsection{Algorithmic Generation of Graphs}

A convenient representation of any graph $\Gamma$ 
is provided by an incidence matrix
$I_{\Gamma}$ with matrix elements $I_{\Gamma}(i,j)$, $i,j\in{1\ldots V}$, $V$ 
being the number of vertices of $\Gamma$. 
Let us enumerate the vertices in some way. 
$I_{\Gamma}(i,j)$ is then given by 
\bea
      I_{\Gamma}(i,i) &=& \mbox{number of external lines at $v_i$}\\
      I_{\Gamma}(i,j) &=& \mbox{number of lines connecting vertices $v_i$ and $v_j$},
      \; i,j \in {1\ldots V}~.  
\eea
This representation is not 
unique. For any permutation of vertices, $I_{\Gamma}$ will change unless 
it corresponds to a symmetry of $\Gamma$. 
A possible way 
out of this ambiguity is the introduction of a complete order relation
among the $I$s. 
(For instance, $I^{(1)}>I^{(2)}$ if $(i^\prime,j^\prime)$ exists such that 
$I^{(1)}(i^\prime,j^\prime)>I^{(2)}(i^\prime,j^\prime)$,
and $I^{(1)}(i,j)=I^{(2)}(i,j)$ for all $i=i^\prime$, $j<j^\prime$ and for all
$i<i^\prime$ and arbitrary $j$.)
A unique representation of the graph is then defined as the maximum over all
permutations of the vertices of $\Gamma$,
i.e. as $I_\Gamma^{\rm max} = \max_{\pi\in P_V} I^{\pi}_{\Gamma}$,
where
$I_\Gamma^\pi(i,j) = I_\Gamma(\pi(i),\pi(j))$.
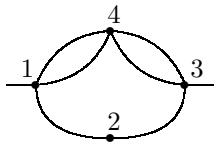
\begin{figure}[h]
\caption{\label{inci} Example for a graph with 2 external and 6 internal lines
with labels attached to the vertices.}

\begin{center}
\setlength{\unitlength}{0.8cm}
\begin{picture}(10.0,2.0)

\put(3.0,0.0){
\setlength{\unitlength}{0.55cm}
\begin{picture}(4.0,0.70)

{\linethickness{0.3pt}
\qbezier(0.100,0.800)(0.100,-0.500)(1.900,-0.500)
\qbezier(3.700,0.800)(3.700,-0.500)(1.900,-0.500)
\qbezier(0.100,0.8000)(0.6000,2.02000)(1.9000,2.1000)
\qbezier(0.100,0.8000)(1.4000,0.88000)(1.9000,2.1000)
\qbezier(3.700,0.8000)(3.2000,2.02000)(1.9000,2.100)
\qbezier(3.700,0.8000)(2.4000,0.88000)(1.9000,2.100)

}

\put(-0.2000,1.2000){\makebox(0.2,0){$1$}}
\put(0.1000,0.8000){\circle*{0.16}}
\put(3.9000,1.2000){\makebox(0.2,0){$3$}}
\put(3.7000,0.8000){\circle*{0.16}}

\put(1.9000,2.5000){\makebox(0.2,0){$4$}}
\put(1.9000,2.1000){\circle*{0.16}}

\put(1.9000,-0.1000){\makebox(0.2,0){$2$}}
\put(1.9000,-0.5000){\circle*{0.16}}
 
\put(0.1000,0.8000){\line(-1,0){0.7}}
\put(3.7000,0.8000){\line(1,0){0.7}}

\end{picture}
}

\end{picture}
\end{center}

\end{figure}
For the simple example of the graph of
Fig.~\ref{inci}
$I_\Gamma$ and $I^{\rm max}_{\Gamma}$ 
are given by
\be
   I_\Gamma \; = \; \left( 
   \begin{array}{ cccc }
    1 & 1 & 0 & 2  \\
      & 0 & 1 & 0  \\
      &   & 1 & 2  \\
   \multicolumn{2}{c}{\mbox{$\cdots$}} &  & 0  \\
   \end{array} \right)
   \quad , \quad
    I_\Gamma^{\rm max} \; = \; \left( 
   \begin{array}{ cccc }
    1 & 2 & 1 & 0  \\
      & 0 & 0 & 2  \\
      &   & 0 & 1  \\
   \multicolumn{2}{c}{\mbox{$\cdots$}} &  & 1  \\
   \end{array} \right).
\ee

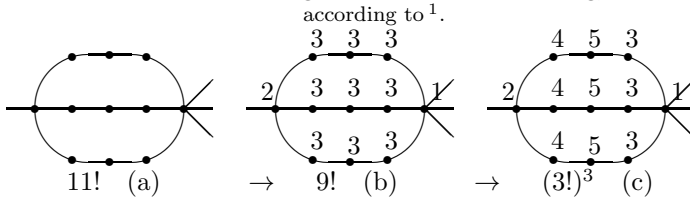
\begin{figure}[h]
\caption{\protect\label{preorder} Different levels of ordering of vertices:
a. no ordering, b. according to \protect\cite{luescher},
c. according to \protect\cite{reisz}.}

\begin{center}
\setlength{\unitlength}{0.8cm}
\begin{picture}(10.0,2.10)

\put(-1.0,0.300){
\setlength{\unitlength}{0.55cm}
\begin{picture}(4.0,0.7)

\put(1.9000,0.8000){\oval(3.600,2.600)}
\put(0.1000,0.8000){\circle*{0.16}}
\put(3.7000,0.8000){\circle*{0.16}}

\put(1.0000,2.05000){\circle*{0.16}}
\put(1.9000,2.1000){\circle*{0.16}}
\put(2.8000,2.05000){\circle*{0.16}}

\put(1.0000,-0.4500){\circle*{0.16}}
\put(1.9000,-0.5000){\circle*{0.16}}
\put(2.8000,-0.4500){\circle*{0.16}}
\put(0.1000,0.8000){\line(1,0){3.6000}}
\put(1.0000,0.8000){\circle*{0.16}}
\put(1.9000,0.8000){\circle*{0.16}}
\put(2.8000,0.8000){\circle*{0.16}}
\put(0.1000,0.8000){\line(-1,0){0.7}}
\put(3.7000,0.8000){\line(1,0){0.7}}
\put(3.7000,0.8000){\line(1,1){0.7}}
\put(3.7000,0.8000){\line(1,-1){0.7}}

\put(1.4,-1.0){\makebox(1.2,0){$11! \quad$(a)}}

\end{picture}
}

\put(3.0,0.300){
\setlength{\unitlength}{0.55cm}
\begin{picture}(4.0,0.7)
\put(1.9000,0.8000){\oval(3.600,2.600)}
\put(-0.2000,1.2000){\makebox(0.2,0){$2$}}
\put(0.1000,0.8000){\circle*{0.16}}
\put(3.9000,1.2000){\makebox(0.2,0){$1$}}
\put(3.7000,0.8000){\circle*{0.16}}

\put(1.0000,2.5000){\makebox(0.2,0){$3$}}
\put(1.0000,2.05000){\circle*{0.16}}
\put(1.9000,2.5000){\makebox(0.2,0){$3$}}
\put(1.9000,2.1000){\circle*{0.16}}
\put(2.8000,2.5000){\makebox(0.2,0){$3$}}
\put(2.8000,2.05000){\circle*{0.16}}

\put(1.0000,0.0300){\makebox(0.2,0){$3$}}
\put(1.0000,-0.4500){\circle*{0.16}}
\put(1.9000,-0.1000){\makebox(0.2,0){$3$}}
\put(1.9000,-0.5000){\circle*{0.16}}
\put(2.8000,0.0300){\makebox(0.2,0){$3$}}
\put(2.8000,-0.4500){\circle*{0.16}}
\put(0.1000,0.8000){\line(1,0){3.6000}}
\put(1.0000,1.3000){\makebox(0.2,0){$3$}}
\put(1.0000,0.8000){\circle*{0.16}}
\put(1.9000,1.3000){\makebox(0.2,0){$3$}}
\put(1.9000,0.8000){\circle*{0.16}}
\put(2.8000,1.3000){\makebox(0.2,0){$3$}}
\put(2.8000,0.8000){\circle*{0.16}}
\put(0.1000,0.8000){\line(-1,0){0.7}}
\put(3.7000,0.8000){\line(1,0){0.7}}
\put(3.7000,0.8000){\line(1,1){0.7}}
\put(3.7000,0.8000){\line(1,-1){0.7}}

\put(-0.50,-1.00){\makebox(3.50,0.00){$\to \quad\; 9! \quad$(b)}}

\end{picture}
}

\put(7.0,0.300){
\setlength{\unitlength}{0.55cm}
\begin{picture}(4.0,0.7)
\put(1.9000,0.8000){\oval(3.600,2.600)}
\put(-0.2000,1.2000){\makebox(0.2,0){$2$}}
\put(0.1000,0.8000){\circle*{0.16}}
\put(3.9000,1.2000){\makebox(0.2,0){$1$}}
\put(3.7000,0.8000){\circle*{0.16}}

\put(1.0000,2.5000){\makebox(0.2,0){$4$}}
\put(1.0000,2.05000){\circle*{0.16}}
\put(1.9000,2.5000){\makebox(0.2,0){$5$}}
\put(1.9000,2.1000){\circle*{0.16}}
\put(2.8000,2.5000){\makebox(0.2,0){$3$}}
\put(2.8000,2.05000){\circle*{0.16}}

\put(1.0000,0.0300){\makebox(0.2,0){$4$}}
\put(1.0000,-0.4500){\circle*{0.16}}
\put(1.9000,-0.1000){\makebox(0.2,0){$5$}}
\put(1.9000,-0.5000){\circle*{0.16}}
\put(2.8000,0.0300){\makebox(0.2,0){$3$}}
\put(2.8000,-0.4500){\circle*{0.16}}
\put(0.1000,0.8000){\line(1,0){3.6000}}
\put(1.0000,1.3000){\makebox(0.2,0){$4$}}
\put(1.0000,0.8000){\circle*{0.16}}
\put(1.9000,1.3000){\makebox(0.2,0){$5$}}
\put(1.9000,0.8000){\circle*{0.16}}
\put(2.8000,1.3000){\makebox(0.2,0){$3$}}
\put(2.8000,0.8000){\circle*{0.16}}
\put(0.1000,0.8000){\line(-1,0){0.7}}
\put(3.7000,0.8000){\line(1,0){0.7}}
\put(3.7000,0.8000){\line(1,1){0.7}}
\put(3.7000,0.8000){\line(1,-1){0.7}}

\put(-0.5,-01.0){\makebox(3.50,0.00){$\to \quad\; (3!)^3 \quad$(c)}}

\end{picture}
}

\end{picture}
\end{center}

\end{figure}

\noindent
A CPU-time consuming part in computer aided LCEs is the 
complete and unique generation of graphs. 
It is performed iteratively with a certain algorithm
for adding new lines and new vertices. It
turns out that in order to make the algorithm both complete (i.e. generating
{\it all} graphs that contribute) and efficient, 
multiple generations cannot be completely avoided.
Let us assume that for a particular class of graphs
we have just generated the $n+1st$ graph after the 
first topologically inequivalent $n$'s have been generated.  
The time consuming procedure is then to find out whether the $n+1st$ 
graph already exists among the first $n$s or not. 
One has to run through all $I_{\Gamma_i}, i=1,\ldots ,n$
to decide whether $\Gamma_{n+1}$ is topologically inequivalent 
to each of the first $n$s. 
A comparision between any 2 graphs is cheap
if they are uniquely represented. 
If it is achieved via $I^{\rm max}_{\Gamma}$ 
as indicated above, 
$V!$ permutations of vertices have to be performed for every newly generated
graph to obtain this representation.
This is quite a large 
number if one goes to a high order in the expansion parameter (for example it 
is $11!$ for the graph of order $\kappa^{12}$ with $12$ internal lines
shown in 
Fig.\ref{preorder}.) 

To avoid the high factorials of permutations of incidence matrices, one needs 
a refined representation of graphs. One possibility is to first 
(partially) order the {\it vertices} and then to perform a maximization procedure. 
A vertex ordering that takes all {\it local} properties 
of vertices into account has been introduced by 
L\"uscher and Weisz \cite{luescher}. With this improvement the 14th order 
in $\kappa$ became feasible.
A further refinement was an extended ordering
proposed in \cite{reisz} which in addition accounts for nonlocal topological
properties to distinguish vertices in combination with an
iterated ordering. The type
of ordering of vertices (if there is any) is reflected 
in the labelling of the vertices, cf. Fig.~\ref{preorder}.


\vskip10pt
Fig.~\ref{preorder}~a shows a graph $\Gamma$ with no labelling 
corresponding to no ordering, Fig.~\ref{preorder}~b with a 
ordering according to \cite{luescher} and Fig.~\ref{preorder}~c 
with an extended ordering according to \cite{reisz}.
The advantage now is that in order to define a final unique 
representation by means
of the above maximization procedure, one has to perform permutations 
only among vertices with equal labels, this is $11!$ for
Fig.~\ref{preorder}~a, 
$9!$ for Fig.~\ref{preorder}~b, but only $3!^3$ for Fig.~\ref{preorder}~c.
At best the number of remaining permutations is as small as the symmetry 
factor of the graph, i.e. the number of simultaneous permutations of rows 
and columns that leave $I_{\Gamma}$ invariant.
It turned out that the gain in CPU-time that is achieved 
because of the reduced number of permutations in $I$ is considerably 
larger than the loose because of the ordering of vertices. 
Further details on the (extended) ordering and the
algorithmic performance can be found in \cite{reisz}.


\section{The Monotony Criterion and its Application to Monte Carlo
Simulations}
 

In LCEs the action is split into a sum of ultralocal parts
$\stackrel{\circ}{S}$ 
and a next neighbour part $S_{nn}$ with
next neighbour couplings $\propto \kappa$. 
A Taylor expansion in $\kappa$ of the logarithm of the partition
function $\ln Z$ about the ultralocal contribution to $\ln Z$ finally leads 
to graphical expansions of n-point susceptibilities $\chi_n$ with coefficients
$a_\mu^{(n)}$.
For every $\mu$, $n$, $a_\mu^{(n)}$ is a sum over all 
connected graphs 
with $\mu$ internal and $n$ external lines each of which adds as its weight
a product of the
inverse topological symmetry factor, an internal symmetry factor, a
lattice embedding factor, and a product of vertex contributions 
depending on the couplings involved in $\stackrel{\circ}{S}$. It is only the
embedding factor that depends on the topology of the particular
lattice that will change in passing from an infinite to a finite
volume.

\vskip10pt

For a certain interval of the scaling region response
functions  with 
a nonanalytic behaviour in the infinite volume limit show different
monotony behaviour for 1st and 2nd order transitions.
Examples for such functions are the specific heat and order parameter
susceptibilities. They are
increasing in volume in a certain neighbourhood of $T_c$ for 2nd order
transitions, and decreasing for 1st order transitions for some
range in the scaling region, which has to be further specified. 
For definiteness we fix the notation in terms of order parameter
susceptibilities $\chi$. 
In particular,
at $T_c(V)$ , $\chi$ has a "$\delta$-function" or power law type of
singularity for a 1st or 2nd order transition in the thermodynamic limit,
respectively. It is this difference that is responsible for
the different monotony properties 
in the finite volume.
In \cite{hilde} we have made these statements more precise in order to
show that the monotony behaviour of the susceptibilities is neither a 
pecularity of specific models nor an artifact of the series expansion. It is 
a generic feature of models with first and second order transitions if 
the standard assumptions on their finite size scaling behaviour apply. Here
we further comment on applications to Monte Carlo calculations.

Choose two volumes $V_1$, $V_2$ both sufficiently large in order to 
satisfy the 
standard assumptions of a FSS
so that the regular contribution to $\chi$ and its induced 
generic volume dependence are neglegible.
Let $V_1\ll V_2$ as argued below. Let $\lambda$
denote a generic coupling parametrizing the transition surface in 
coupling constant space. Choose 
$\delta$ such that
$c_1 \sigma(V_2)^{\frac{1}{1+\epsilon}} < |\delta| < 
  c_2 \sigma(V_1)$,
where $c_1$, $c_2$ and $\epsilon$ are positive constants and $\sigma$
denotes the width of the critical region.
Define 
\be \label{fss.rml}
   r_{V_1,V_2} := 1 - \frac{ \chi_2(\delta + \kappa_c(V_1),V_1)}
    {\chi_2(\delta+\kappa_c(V_2),V_2)}  .
\ee
The monotony criterion says that
\be \label{cde}
    r_{V_1,V_2} \; \left\{ 
    \begin{array}{r@{\; ,\;} l }
    > 0 & \mbox{2nd order} \\
    < 0 & \mbox{1st order }\\
    = 0 & \mbox{tricritical point for 
         $\partial r/ \partial \lambda \not=0$.}
    \end{array} \right. 
\ee
The difference of using the monotony criterion in Monte Carlo simulations or in
series representations of $\chi$ comes from the second volume $V_2>V_1$
that must be finite in Monte Carlo simulations, but can be infinite in LCEs.
For $V_2<\infty$ also in case of 1st order transitions a small neighbourhood 
around the peak of $\chi$ exists where $\chi$ increases with $V$ because of the
rounding of the $\delta$-singularity. Thus $V_2$ should be chosen sufficiently
larger than $V_1$ so that the widths of the critical regions behave like
$\sigma(V_2)\ll \sigma(V_1)$. 
For finite $\sigma(V_2)$ the points $T$ or $\kappa$ 
at which the
$\chi$s are evaluated should be chosen at the same distance from the 
(volume dependent) position of the peak of $\chi$.

Deviations from these predictons in Monte Carlo calculations may be caused by
$\bullet$ $V_1,V_2$ both too small
$\bullet$ $V_2/V_1$ too small
$\bullet$ $(\kappa-\kappa_c(V))/\kappa_c(V)$ too large, i.e. contributions 
from the regular part cannot be neglected.
$\bullet$ $(\kappa-\kappa_c(V))/\kappa_c(V)$ too small, 
so that $r>0$ even for 1st order.
 
In particular the interval of allowed $\kappa$s in Eq.(\ref{cde})
depends on the coupling $\lambda$. 
Note that the ratios of Eq. (\ref{cde}) do not include the generic
volume dependence that is induced by the analytic part of $\chi$.
Assuming that one is lucky in simultaneously matching these constraints,
a result of $r_{V_1,V_2}>(<)0$ for all $\kappa$ from the specified 
scaling region
excludes (indicates) a 1st order transition, respectively. 

If there is a crossover phenomenon rather 
than a true phase transition, $r$ should vanish 
for a finite neighbourhood of $\lambda$
for sufficiently large
$V_1,V_2$. 

An application of the monotony criterion to LCEs in a scalar $O(N)$ theory 
with $\Phi^4$ and $\Phi^6$-terms
in 3 dimensions with $N=1$ or $N=4$ improved the localizability of the
tricritical line in this model by 2 orders of magnitude \cite{hilde}.


\section{Dynamical Linked Cluster Expansions}


As {\it Dynamical Linked Cluster Expansions} (DLCEs) we call LCEs with
dynamical next neighbour couplings. The next neighbour couplings still
play the role of expansion
parameters, but are endowed with their own dynamics. More precisely DLCEs
amount to the following generalization of the familiar linked cluster 
expansions. 

Let $\stackrel{\circ}{\Lambda}$ denote the set of sites of a 
hypercubic lattice,
$\stackrel{1}{\Lambda}$ the set of unordered pairs of sites
$l=(xy)$, not necessarily nearest neighbours. We consider 
a statistical system characterized by the partition function
\be
Z(H,J;v)={\cal{N}} \; \int \prod_{x\in\stackrel{\circ}{\Lambda}} d\Phi_x \cdot
 \prod_{l\in\stackrel{1}{\Lambda}} d U_l \cdot
 \exp{[ - S(\Phi,U;v) + \sum_{x\in\stackrel{\circ}{\Lambda}} H_x \Phi_x
  + \sum_{l\in\stackrel{1}{\Lambda}} J_l U_l ]} 
\ee
with normalization factor ${\cal{N}}$ such that $Z(0,0,v)=1$ and an action $S$
having the form
\be \label{action}
S(\Phi,U;v)=\sum_{x\in\stackrel{\circ}{\Lambda}} 
            \stackrel{\circ}{S}(\Phi_x) 
            +\sum_{l\in\stackrel{1}{\Lambda}} \stackrel{1}{S}(U_l)
              -\frac{1}{2} \sum_{x,y\in\stackrel{\circ}{\Lambda}}
            v_{xy} \Phi_x U_{(xy)} \Phi_y
\ee
with $v_{xy}=v_{yx}$, $v_{xx}=0$.
$\Phi$ can be
an $N$-component scalar field associated with the lattice sites $x$
or a $Z_N$-spin, likely $U$ can be any
scalar, vector or tensor field associated with lattice links $l$.
The action $S$ is split into two
ultralocal parts $\stackrel{\circ}{S}$ and $\stackrel{1}{S}$, depending 
on single sites and on
single links via the fields $\Phi_x$ and $U_l$, and an interaction
part with coupling constants $v_{xy}$. 
Note that formally the former coupling $v_{xy}$ is
replaced by $v_{xy} \cdot U_{xy}$ with the dynamics of $U$ governed by
$\stackrel{1}{S}$.

A new type of graphical expansion of $n$-point correlation functions
is then induced by
Taylor expanding $\ln{Z(H,J;v)} \equiv W(H,J;v)$ about $v=0$ with
\be
 W(H,J;v=0) = \sum_{x\in\stackrel{\circ}{\Lambda}} \stackrel{\circ}{W}(H)
          + \sum_{l\in\stackrel{1}{\Lambda}} \stackrel{1}{W}(J).
\ee
To illustrate the new features, we list the terms occurring in  
the second derivatives 
in the Taylor expansion of $W$. Written as indices, let $J_{(lm)}$ and
$H_l$ stand for the derivatives of $W$ with respect
to $J_{(lm)}$ and $H_l$ at $v=0$, respectively. The expressions
\bea
&& W_{H_j} \cdot W_{J_{ij}} \cdot W_{H_l~H_i} \cdot W_{H_m} \cdot W_{J_{lm}} 
\nonumber \\
&& W_{J_{ij}} \cdot W_{H_l~H_i} \cdot W_{H_m~H_j} \cdot W_{J_{lm}} 
\nonumber \\
&& W_{H_l~H_i} \cdot W_{H_m~H_j} \cdot W_{J_{lm}~J_{ij}} 
\label{ww} \\
&& W_{H_j} \cdot W_{H_l~H_i} \cdot W_{H_m} \cdot W_{J_{lm}~J_{ij}}
\nonumber \\
&& W_{H_i} \cdot W_{H_j} \cdot W_{H_l} \cdot W_{H_m} \cdot W_{J_{lm}~J_{ij}}
\nonumber
\eea
suggest a graphical representation according to 
Fig. \ref{rule1},
respectively.


\begin{figure}[h]
\caption{\label{rule1} Graphical representation of the terms of Eq.(~\ref{ww}), 
respectively.}

\begin{center}
\setlength{\unitlength}{0.8cm}
\begin{picture}(10.0,2.0)

\put(1.0,0.50){
\setlength{\unitlength}{0.55cm}
\begin{picture}(0.50,2.0)

\put(0.000,0.0000){\circle*{0.16}}
\put(2.000,0.0000){\circle*{0.16}}
\put(1.000,2.0000){\circle*{0.16}}
\put(0.000,0.0000){\line(1,2){1.0}}
\put(2.000,0.0000){\line(-1,2){1.0}}

\end{picture}
}

\put(3.0,0.50){
\setlength{\unitlength}{0.55cm}
\begin{picture}(0.50,2.0)

\put(1.000,0.0000){\circle*{0.16}}
\put(1.000,2.0000){\circle*{0.16}}
{\linethickness{0.3pt}
\qbezier(1.000,0.0000)(0.000,1.000)(1.000,2.000)
\qbezier(1.000,0.0000)(2.000,1.000)(1.000,2.000)
}
 
\end{picture}
}

\put(5.0,0.50){
\setlength{\unitlength}{0.55cm}
\begin{picture}(0.50,2.0)

\put(1.000,0.0000){\circle*{0.16}}
\put(1.000,2.0000){\circle*{0.16}}
{\linethickness{0.3pt}
\qbezier(1.000,0.0000)(0.000,1.000)(1.000,2.000)
\qbezier(1.000,0.0000)(2.000,1.000)(1.000,2.000)
}
{\thinlines
\put(-0.150,1.0000){\line(1,0){0.3}}
\put(0.3500,1.0000){\line(1,0){0.3}}
\put(0.8500,1.0000){\line(1,0){0.3}}
\put(1.3500,1.0000){\line(1,0){0.3}}
\put(1.8500,1.0000){\line(1,0){0.3}}
}
 
\end{picture}
}

\put(7.0,0.50){
\setlength{\unitlength}{0.55cm}
\begin{picture}(0.50,2.0)

\put(0.000,0.0000){\circle*{0.16}}
\put(2.000,0.0000){\circle*{0.16}}
\put(1.000,2.0000){\circle*{0.16}}
\put(0.000,0.0000){\line(1,2){1.0}}
\put(2.000,0.0000){\line(-1,2){1.0}}
{\thinlines
\put(-0.150,0.9000){\line(1,0){0.3}}
\put(0.3500,0.9000){\line(1,0){0.3}}
\put(0.8500,0.9000){\line(1,0){0.3}}
\put(1.3500,0.9000){\line(1,0){0.3}}
\put(1.850,0.9000){\line(1,0){0.3}}
}

\end{picture}
}

\put(9.0,0.50){
\setlength{\unitlength}{0.55cm}
\begin{picture}(0.50,2.0)

\put(0.200,0.0000){\circle*{0.16}}
\put(0.200,2.0000){\circle*{0.16}}
\put(1.800,0.0000){\circle*{0.16}}
\put(1.800,2.0000){\circle*{0.16}}
\put(0.200,0.0000){\line(0,1){2.0}}
\put(1.800,0.0000){\line(0,1){2.0}}
{\thinlines
\put(-0.150,1.0000){\line(1,0){0.3}}
\put(0.3500,1.0000){\line(1,0){0.3}}
\put(0.8500,1.0000){\line(1,0){0.3}}
\put(1.3500,1.0000){\line(1,0){0.3}}
\put(1.8500,1.0000){\line(1,0){0.3}}
}

\end{picture}
}

\end{picture}
\end{center}

\end{figure}
The dashed lines
indicate the new type of local connectivity, originating 
from $\stackrel{1}{W}$, 
n lines are now connected if at least one of the following conditions is met.
$\bullet$ The lines share a common vertex $\stackrel{\circ}{v}_n^c \equiv
\frac{\partial^n\stackrel{\circ}{W}}{\partial H^n}$ as before.
$\bullet$ The lines are part of a multiple-line sharing a common factor 
$\stackrel{1}{g}_n^c \equiv
\frac{\partial^n\stackrel{1}{W}}{\partial J^n}$.

Thus the building blocks of the graphical expansion for DLCEs are
m-point vertices
and n-multiple-lines as shown in Fig.~\ref{rule2}.


\begin{figure}[h]
\caption{\label{rule2} Graphical rules for DLCES.}

\begin{center}
\setlength{\unitlength}{0.8cm}
\begin{picture}(10.0,2.50)

\put(-2.0,0.300){
\setlength{\unitlength}{0.55cm}
\begin{picture}(0.50,2.0)

\put(0.200,0.0000){\circle*{0.16}}
\put(1.800,0.0000){\circle*{0.16}}
\put(5.000,0.0000){\circle*{0.16}}

\put(2.8,1.3){\makebox(0.1,0.00){$\cdot$}}
\put(3.4,1.3){\makebox(0.1,0.00){$\cdot$}}
\put(4.0,1.3){\makebox(0.1,0.00){$\cdot$}}

\put(0.200,2.0000){\circle*{0.16}}
\put(1.800,2.0000){\circle*{0.16}}
\put(5.000,2.0000){\circle*{0.16}}
\put(0.200,0.0000){\line(0,1){2.0}}
\put(1.800,0.0000){\line(0,1){2.0}}
\put(5.000,0.0000){\line(0,1){2.0}}

\put(0.2,-0.5){\makebox(0.1,0.00){$1$}}
\put(1.8,-0.5){\makebox(0.1,0.00){$2$}}
\put(5.0,-0.5){\makebox(0.1,0.00){$n$}}

{\thinlines
\put(-0.150,0.9000){\line(1,0){0.3}}
\put(0.3500,0.9000){\line(1,0){0.3}}
\put(0.8500,0.9000){\line(1,0){0.3}}
\put(1.3500,0.9000){\line(1,0){0.3}}
\put(1.8500,0.9000){\line(1,0){0.3}}
\put(2.3500,0.9000){\line(1,0){0.3}}
\put(2.8500,0.9000){\line(1,0){0.3}}
\put(3.3500,0.9000){\line(1,0){0.3}}
\put(3.8500,0.9000){\line(1,0){0.3}}
\put(4.3500,0.9000){\line(1,0){0.3}}
\put(4.85500,0.9000){\line(1,0){0.3}}
}

\end{picture}
}

\put(3.0,0.300){
\setlength{\unitlength}{0.55cm}
\begin{picture}(0.50,2.0)

{\Large
\put(0.0,1.0){\makebox(0.1,0.00){$=$}}
\put(2.0,1.0){\makebox(1.0,0.00)
{$\frac{\partial^n \stackrel{1}{W}(J)}{\partial J^n}$}}
}

\end{picture}
}

\put(7.0,0.80){
\setlength{\unitlength}{0.55cm}
\begin{picture}(0.50,2.0)

\put(2.000,0.0000){\circle*{0.16}}

\put(2.0000,0.0000){\line(-1,2){1.0}}
\put(2.0000,0.0000){\line(-1,1){1.0}}
\put(2.0000,0.0000){\line(-1,-2){1.0}}
 
\put(0.6,0.0){\makebox(0.1,0.00){$\cdot$}}
\put(0.6,-0.5){\makebox(0.1,0.00){$\cdot$}}
\put(0.6,-1.0){\makebox(0.1,0.00){$\cdot$}}

\put(0.6,2.1){\makebox(0.1,0.00){$1$}}
\put(0.6,1.2){\makebox(0.1,0.00){$2$}}
\put(0.6,-2.1){\makebox(0.1,0.00){$n$}}

\end{picture}
}

\put(9.0,0.80){
\setlength{\unitlength}{0.55cm}
\begin{picture}(0.50,2.0)

{\Large
\put(0.0,0.0){\makebox(0.1,0.00){$=$}}
\put(2.0,0.0){\makebox(1.0,0.00)
{$\frac{\partial^n \stackrel{0}{W}(H)}{\partial H^n}$}}
}

\end{picture}
}

\end{picture}
\end{center}

\end{figure}


\vskip10pt
Note that the graphs of the former LCEs are contained as a small
subclass of the graphs of DLCEs. If we replace for constant $U$ $vU$
by $v$ and choose $\stackrel{1}{W}(J)$ as $J_l\cdot v$, we have
$\frac{\partial W}{\partial J}=v$ (1-lines
as before), but $\frac{\partial^n\stackrel{1}{W}}{\partial J^n}=0$
for $n>1$ (corresponding to no multiple-lines).

A rapid increase in the number of graphs contributing to an n-point 
susceptibility sets in at low order in the expansion  in $\kappa$. For
example the number of graphs contributing to $\chi_2$ to order $\kappa^4$
is of the order of $100$, while it is by an order of magnitude smaller
in the pure LCE of $\chi_2$.

The generic form for the action in Eq.~(\ref{action}) has interesting physical 
applications to systems with coupled dynamics with (fast) spins
and (slow) interactions in spin glasses
and neural networks, in which both spins 
and couplings are endowed with a dynamical law and evolve in time 
\cite{sherrington}.
 
So far the graphs for LCEs could be generated on a SUN-workstation
with a CPU-time of the order of days and a working space of the order
of $100$~MB. Work on computer aided DLCEs is in progress \cite{tom}.


\end{document}